# Half-solitons in a polariton quantum fluid behave like magnetic monopoles

R. Hivet[1], H. Flayac[2], D. D. Solnyshkov[2], D. Tanese[3], T. Boulier[1], D. Andreoli[1], E. Giacobino[1], J. Bloch[3], A. Bramati[1*], G. Malpuech[2], A. Amo[3*]

[1] *Laboratoire Kastler Brossel, Université Pierre et Marie Curie, Ecole Normale Supérieure et CNRS, UPMC case 74, 4 place Jussieu, 75005 Paris, France*

[2] *Institut Pascal, PHOTON-N2, Clermont Université, University Blaise Pascal, CNRS, 24 avenue des Landais, 63177 Aubière cedex, France*

[3] *Laboratoire de Photonique et Nanostructures, CNRS, Route de Nozay, 91460 Marcoussis, France*

* e-mail address : bramati@spectro.jussieu.fr; alberto.amo@lpn.cnrs.fr .

**Monopoles are magnetic charges, point-like sources of magnetic field. Contrary to electric charges they are absent in Maxwell's equations and have never been observed as fundamental particles. Quantum fluids such as spinor Bose-Einstein condensates have been predicted to show monopoles in the form of excitations combining phase and spin topologies. Thanks to its unique spin structure and the direct optical control of the fluid wavefunction, an ideal system to experimentally explore this phenomenon is a condensate of exciton-polaritons in a semiconductor microcavity. We use this system to create half-solitons, non-linear excitations with mixed spin-phase geometry. By tracking their trajectory, we demonstrate that half-solitons behave as monopoles, magnetic charges accelerated along an effective magnetic field present in the microcavity. The field-induced spatial separation of half-solitons of opposite charges opens the way to the generation of magnetic currents in a quantum fluid.**

Magnetic monopoles are the magnetic counterparts of electric charges, characterised by a divergent field. Though they have never been observed in the form of elementary particles, the seminal work of Dirac[1] showed that monopoles are allowed by the laws of quantum mechanics. The elusiveness of their observation has motivated the study of monopole analogues in the form of quasiparticles in many-body systems, and they have been recently observed in spin ice crystals[2-4]. Other systems where magnetic monopoles have been predicted are spinor Bose-Einstein condensates, which present quantum properties at the macroscopic scale, and show fascinating spin phenomena thanks to the vectorial character of the order parameter[5-7]. Two-component spinor condensates can be described by a global phase and a polarisation angle, the latter representing the relative phase between the two spin populations. Compared to the case of scalar condensates [8], this feature enlarges substantially the space of possible excitations, as they can now combine both phase and polarisation topologies[9,10]. One example are spin-phase topological defects arranged in the form of vectorial vortices and solitons[11,12]. The arrangement can be done in such a way that the overall phase winding of the wavefunction around the defect is stored both in the polarisation angle and the wavefunction phase, giving rise to a divergent spin field analogous to the electric field generated by a point charge[13-15]. A vortex or a soliton with such a spin-phase texture would be sensitive to a magnetic field, and it would behave like a magnetic charge (a monopole), accelerated along an applied magnetic field with a direction defined by the sign of the charge.

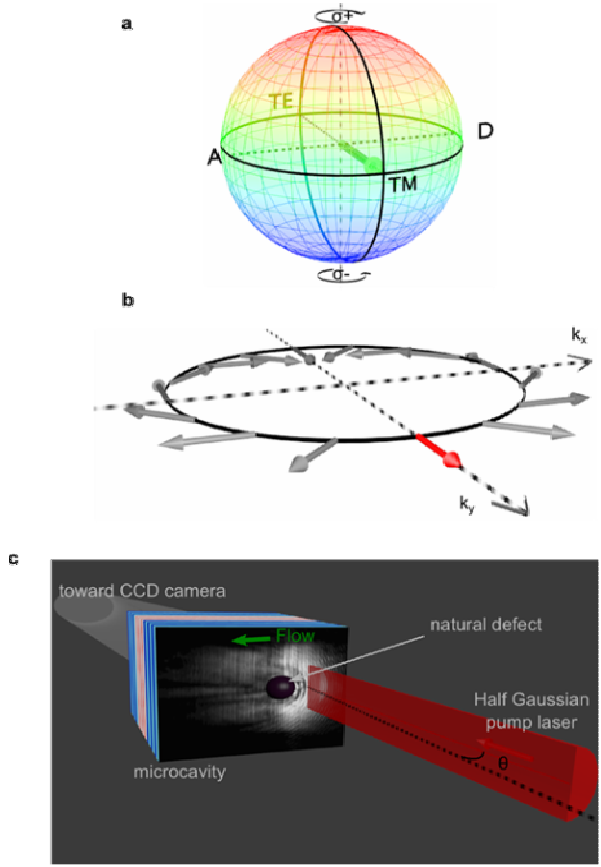

**Figure 1. Polariton pseudospin, effective magnetic field and experimental setup**. **a** Bloch sphere representing all the possible spin configuration of the polariton gas and the associated polarisations. **b** Direction of the effective magnetic field created by the TE-TM splitting for polaritons propagating in different directions. **c** Scheme of the resonant injection of the polariton gas above a round potential barrier present in the sample. Half-solitons nucleate in its wake.

The observation of these objects remains experimentally elusive in atomic spinor condensates. The difficulty in generating such spin-phase structures resides in the required fine manipulation of the local spin and phase[16], and on the need of significant antiferromagnetic interactions to stabilise them[14,15]. In this sense, exciton-polariton condensates currently appear as a well suited system to evidence and study such original effects in quantum fluids. On the one hand, polariton fluids are easy to manipulate with standard optical techniques[17-20]. On the other hand, they possess two possible spin projections and strongly spin-dependent particle interactions of the antiferromagnetic type[21,22], allowing the formation and stabilisation of non-linear spin excitations.

One kind of spin-phase topological defects already reported in polariton quantum fluids are the so-called half-vortices[23,24]. Different from integer quantized vortices in scalar fluids where the phase winds from *zero* to $2\pi$ when going around the vortex core[25], half-vortices present a simultaneous rotation of $\pi$ of both the phase and the polarisation angle around their core. These objects have been recently predicted to behave like monopoles[26], but experiments have so far reported half-vortices pinned to local inhomogeneities of the sample[24], preventing any probing of the monopole physics.

In this work we report the generation of a different kind of vectorial topological excitation in a flowing polariton condensate, oblique dark half-solitons. They are characterised by a notch in the polariton density of the fluid, and a simultaneous phase and polarisation rotation of $\pi/2$ in the condensate wavefunction across the soliton[27] (as opposed to a phase jump of $\pi$ for dark solitons in scalar condensates[28]). This is manifested in the circular polarisation basis as a deep notch present in only one polarisation component. We map the polarisation and phase of these objects evidencing their complex spin structure, and we show that they are indeed accelerated by the action of the intrinsic effective magnetic field present in our microcavities, thus behaving as magnetic monopoles.

Exciton-polaritons are the quasiparticles arising from the strong coupling between excitons and photons confined in semiconductor microcavities[29]. They have recently become a model system for the study of quantum fluid effects such as superfluidity[30], vortex formation[19,25,31] or oblique solitons[28,32]. Their spin structure is especially interesting: polaritons are bosons with only two allowed spin projections $S_z = \pm 1$ on the growth axis of the sample, which couple, respectively, to $\sigma_\pm$ circularly polarised photons in and out of the cavity. Coherent superposition of different spin populations gives rise to polarisation states that can be described by a pseudospin $\vec{S}$ and mapped onto a Bloch sphere (Fig. 1a). The poles represent circular polarisation, the equator linearly polarised states, and intermediate latitudes elliptically polarised states. Another feature of polaritons is the coupling between the spin and motional degrees of freedom (spin-orbit like coupling)[33]. Due to the polarisation-dependent penetration of the electromagnetic field in the distributed mirrors forming the microcavity, the polariton states present a polarization splitting [transverse electric (TE)-transverse magnetic (TM) splitting] resulting in an effective magnetic field $\vec{\Omega}_{TE-TM}$ acting on the polariton pseudospin $\vec{S}$, with strongly momentum-dependent direction and magnitude[33] (Fig. 1b). This field adds a magnetic energy term to the Hamiltonian describing the system of the form[21,22] $H_{TE-TM} = -\vec{S} \cdot \vec{\Omega}_{TE-TM}$ (see Supplementary Material).

In our experiments we inject the polariton fluid in an InGaAs/GaAs/AlGaAs semiconductor microcavity (see Methods) at 10 K by resonant excitation of the lower polariton branch with a *cw* Ti:Sapphire monomode laser. Polarisation-resolved real space images of the polariton fluid in transmission geometry are then recorded on a CCD camera via the photons escaping out of the cavity (Fig. 1c). The fluid is injected at supersonic speed (in-plane momentum of 1.3 μm$^{-1}$, see Methods), upstream from the potential barrier formed by a structural photonic defect present in our sample[28]. In these conditions, as already reported[28,34], a circularly polarised excitation beam leads to the formation of pairs of oblique dark solitons in the wake of the barrier, characterised by a phase jump close to $\pi$ across each notch (see Supplementary Material). In the present experiments, we create a polariton gas with linear polarisation parallel to the flow (TM polarisation, along the *y*-direction). The use of a linearly polarised pump is a key feature to explore the nucleation of spin-phase topological excitations, which can be evidenced by analysing the circularly polarised components of the emission[27].

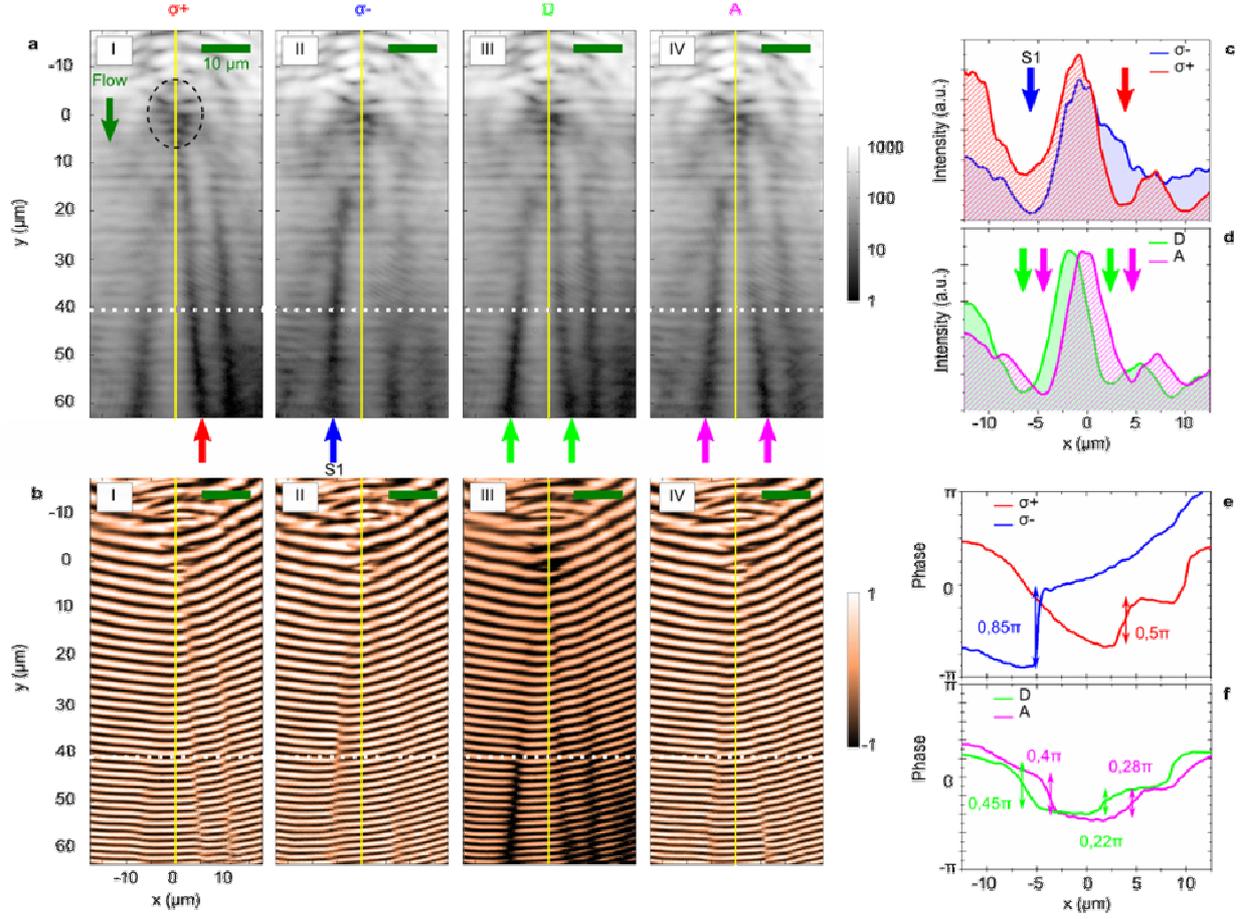

**Figure 2**. **Density and phase tomography of the half-solitons**. **a** Emitted intensity of the polariton gas in $\sigma_+$ and $\sigma_-$ circularly polarised components and in the *diagonal* and *antidiagonal* linearly polarised components with respect to the TM polarisation of injection. Half-solitons spontaneously nucleate in the wake of the potential barrier (dashed circle in **a-I**) and are evidenced as dark traces only present in one circular polarisation component (arrows). The yellow line is a guide to the eye. S1 indicates the half-soliton discussed in the text. **b** Corresponding interferometric images obtained from the interference of the real space emission with a beam of homogeneous phase. **c** Density profiles of the $\sigma_+$ and $\sigma_-$ emission along the dotted line in **a-I** and **a-II**. The arrows indicate the position of the inner half-solitons, only present in a given circular polarisation, with an associated phase jump shown in **e** (obtained from **b-I** and **b-II**). Complementarily, half-solitons appear in the density profiles of both diagonal and antidiagonal polarisations (extracted from the dashed line in **a-III** and **a-IV**), **d**, with phase jumps of half the value measured in circular polarisation, as shown in **f** (obtained from **b-III** and **b-IV**). This characteristic is at the origin of the "half" etymology of half-solitons. Error bars in the estimation of the phase jumps shown in **e** and **f** amount to $\pm 0.08\pi$.

Figure 2a-I shows the formation of two oblique dark solitons to the right of the barrier wake in the $\sigma_+$ component. They can be identified as dark straight notches in the polariton density. These solitons are almost absent in the $\sigma_-$ component (Fig. 2a-II). In turn, in the $\sigma_-$ emission, a deep soliton (S1) clearly appears to the left of the barrier wake (blue arrow), where only a very shallow one is present in $\sigma_+$ (see the profiles in Fig. 2c). The absence of mirror symmetry between Figs. 2a-I and 2a-II arises from the specific form of the potential barrier. The individual dark solitons in each $S_z = \pm 1$ component of the fluid appear as long spatial traces with a high degree of circular polarisation [$\rho_c = (I^+ - I^-)/(I^+ + I^-)$, where $I^\pm$ is the emitted intensity in $\sigma_\pm$ polarisation], as shown in Fig. 3a. Interferometry images obtained by combining the real space emission field with a reference beam of homogeneous phase (Fig. 2b-I and 2b-II) give access to the phase jump across each soliton. For instance, for the soliton S1 observed in $\sigma_-$, we measure a phase jump of $\Delta\theta_- = 0.85\pi$ (Fig. 2e; note that it would be $\pi$ for a perfectly dark soliton), while in the same region the phase in the $\sigma_+$ component does not change ($\Delta\theta_+ \approx 0$).

A dark soliton present in just one spin component of the fluid is the fingerprint of a half-soliton[27]. The mixed spin-phase character of these topological excitations is further evidenced when analysing them in the linear polarisation basis. In the regions where the two circular polarisations are of equal intensity (i.e., the fluid surrounding the half-solitons) we can define a linear polarisation angle $\eta = \frac{\theta_+ - \theta_-}{2}$ and a global phase $\phi = \frac{\theta_+ + \theta_-}{2}$, where $\theta_+$, $\theta_-$ are the local phases of each circularly polarised component[23,27]. In our experiments we directly access the phase jump $\Delta\phi$ and the change of $\eta$ across the solitons by looking at the linearly polarised emission, for instance, in the diagonal and anti-diagonal directions (polarisation plane rotated by +45º and −45º with respect to the TM direction). Figure 2d and 2f show that the half-soliton S1 is also present in these polarisations with a phase jump of $\Delta\phi \approx 0.4\pi$. This confirms that across the half-solitons, $\phi$ undergoes a jump $\Delta\phi \approx 0.85\pi/2 \approx (\Delta\theta_+ + \Delta\theta_-)/2$, that is, *one half* the phase jump observed in the circularly polarised component in which the soliton is present. We also expect a similar jump $\Delta\eta$ of the direction of polarisation. This is illustrated in Fig. 3b, where all the half-solitons present in our fluid (dashed lines extracted from Figs. 2a-I and 2a-II) appear as walls between domains of diagonal (magenta) and anti-diagonal (green) polarisation. Mapping the linear polarisation vector in the vicinity of the soliton S1 (Fig. 4a), we deduce a jump of the polarisation direction of $\Delta\eta \approx 0.32\pi$ (Fig. 4c), close to the ideal expected value of $0.4\pi$.

In the remaining of this letter we will discuss the analogy between half-solitons and magnetic monopoles. In our cavity sample there is an intrinsic effective magnetic field originating from the TE-TM splitting present in the structure[33]. This magnetic field $\vec{\Omega}_{TE-TM}$ points in the direction of the flow (*y*, red arrow in Fig. 1b). As the half-soliton S1 has an inclined trajectory with respect to the flow direction, there is a component of the magnetic field perpendicular to the soliton. This is illustrated in Fig. 4b, which shows the measured degree of circular polarisation together with the linear polarisation direction (equatorial coordinate of the pseudospin in the Bloch sphere). Perpendicular to S1 (dotted line), the pseudospin points away from the half-soliton on both sides. We can calculate the local magnetic energy per unit length as $\int -\vec{S}(x'-x_0)\cdot\vec{\Omega}_{TE-TM}\,dx'$, where the integral is performed along the $x'$ transverse direction, perpendicular to the half-soliton located at $x_0$. The energy has a positive contribution from the left of the half-soliton ($\vec{S}$ and $\vec{\Omega}_{TE-TM}$ pointing in opposite directions), and a negative one from the right ($\vec{S}$ and $\vec{\Omega}_{TE-TM}$ having the same direction). In order for the magnetic energy to be minimised, a magnetic force appears *pushing* the half-soliton towards the left, increasing the negative contribution. As solitons are density notches, their effective mass is negative[8] and, therefore, the acceleration is in the direction opposite to the force. In this way, the half-soliton S1 that appears in the $\sigma_-$ component of the fluid accelerates towards the right, as sketched in Fig. 4b. The direction of the acceleration is opposite for a soliton present in the $\sigma_+$ component (see arrows in Fig. 3b).

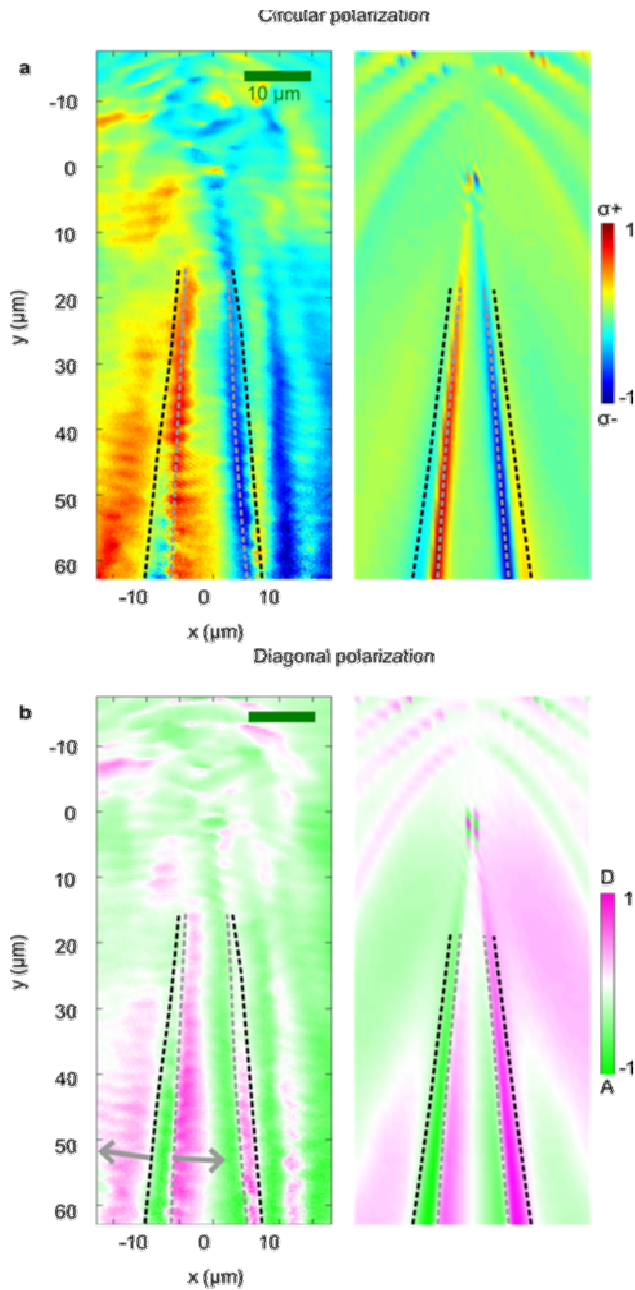

**Figure 3**. **Polarisation texture of half-solitons**. Left panels in **a** and **b** show the measured degree of circular and diagonal polarisations, respectively. The right panels show the calculated patterns from the solution of the non-linear spin dependent Schrödinger equation describing the system in the conditions of the experiment (see Supplementary Material). Dashed lines show the trajectory of the half-solitons extracted from Fig. 1a-I and 1a-II. The trajectories of the half-solitons appear as extrema of circular polarisation and domain walls in diagonal polarisation.

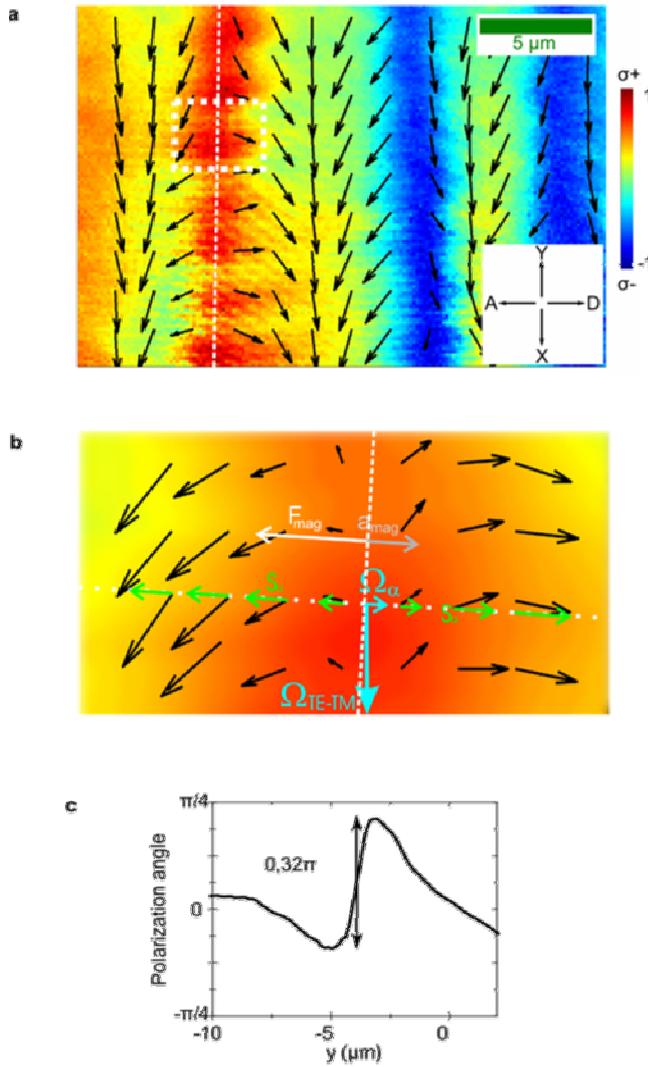

**Figure 4. Magnetic force acting on the half-solitons. a** Complete spin texture of the polariton fluid in the region between 30 and 50 µm below the barrier. The colour scale indicates the degree of circular polarisation (latitude in the Bloch sphere), and the arrows the direction of the linear polarisation as defined in the equator of the Bloch sphere. The half-soliton S1 presents a pseudospin field analogue to that of a point charge in the direction perpendicular to its trajectory, as depicted in **b** for the dash-boxed area of **a**. The polarisation angle jumps from the antidiagonal to the diagonal direction when crossing the soliton, as shown in **c**. The TE-TM effective magnetic field schematized in **b** has a component perpendicular to the half-soliton resulting in its acceleration.

The behaviour we just described is equivalent to that of a magnetic charge in the presence of a magnetic field, that is, a monopole, whose sign is given by that of the circular polarisation in which the half-soliton is present. Indeed, the divergent pseudospin texture along the dotted line in Fig. 4b (green arrows) is the same as the field generated by a point charge in a one-dimensional system. This behaviour allows understanding the mechanisms of formation of the half-solitons. Each soliton nucleated just behind the obstacle does not present a significant degree of circular polarisation, close to be an integer soliton. It can be seen as a superposition of two half-solitons of opposite circularities and, therefore, carrying opposite magnetic charges. The presence of the TE-TM effective magnetic field makes them experience opposite magnetic forces inducing different spatial trajectories depending on their charge, leading to their separation (dashed lines in Fig. 3). The half-solitons pushed towards the centre are slowed down, gain stability and become darker. Those pushed outwards gain velocity and become shallower until they eventually disappear (see Supplementary Material). The trajectories of the external shallow half-solitons are clearly visible in our experiments as domain walls in the diagonal basis, and as maxima of polarisation on the circular basis (black dashed lines in Fig 3). One should however notice that the trajectories of these expelled secondary half-solitons are perturbed far from the obstacle axis by the presence of additional solitons nucleated by the large barrier, particularly on the right side of the images. The monopole behaviour and soliton separation are well reproduced by a non-linear Schrödinger equation including spin and the effective magnetic field present in our microcavities (see Fig. 3 and Supplementary Material).

A remarkable feature of half-soliton monopoles is that they can be thought of as partly photonic charged quasiparticles. Their dynamics can be controlled by applying strain, electric or magnetic fields which would modify the effective magnetic field[35]. Our results inaugurate the study of monopole-analogue physics in quantum fluids, with the great advantages with respect to spin ice crystals that their generation can be well controlled via the phase and density engineering of the polariton wavefunction, and that their trajectories can be easily followed using standard optical techniques. Additionally, thanks to the developed engineering of the polariton landscape[17,18] we open the way to the realisation of magnetronic circuits in a polariton chip.

**Acknowledgements**

We thank R. Houdré for the microcavity sample and P. Voisin for fruitful discussions. This work was supported by the *Agence Nationale de la Recherche* (contract "QUANDYDE"), the RTRA (contract "Boseflow1D"), IFRAF, the FP7 ITNs "Clermont4" (235114) and "Spin-Optronics"(237252), and the FP7 IRSES "Polaphen" (246912). A.B. is a member of the *Institut Universitaire de France*.

**Methods**

**Sample description.** The experimental observations have been performed at 10 K in a $3\lambda/2$ GaAs microcavity containing three $In_{0.05}Ga_{0.95}As$ quantum wells. The top and bottom Bragg mirrors embedding the cavity have, respectively, 21 and 24 pairs of GaAs/AlGaAs alternating layers with an optical thickness of $\lambda/4$, $\lambda$ being the wavelength of the confined cavity mode. The resulting Rabi splitting amounts 5.1 meV, and the polariton lifetime is about 10 ps. During the molecular beam epitaxy growth of the distributed Bragg reflectors of the sample, the slight lattice mismatch between the lattice constants of each layer results in an accumulated stress which relaxes in the form of structural defects. These photonic defects create high potential barriers in the polariton energy landscape.

**Excitation scheme.** To create a polariton fluid we excite the microcavity with a continuous-wave single-mode Ti:Sa laser resonant with the lower polariton branch. We use a confocal excitation scheme in which the laser is focalised in an intermediate plane where a mask is placed in order to hide the upper part of the Gaussian spot. Then, an image of this intermediate plane is created on the sample, producing a spot with the shape of a half Gaussian. Polaritons are resonantly injected in the microcavity with a well-defined wavevector, in the region above the defect. In these conditions, polaritons move out of the excitation spot with a free phase, no more imposed by the pump beam. This is essential for the observation of quantum hydrodynamic effects involving topological excitations with phase discontinuities[28].

The momentum of the injected polaritons is set by the angle of incidence $\alpha$ of the excitation laser on the microcavity. This allows us to control the in-plane wave vector of the polariton fluid through the relation $k = k_0 \sin(\alpha)$, where $k_0$ is the wave vector of the laser field. At the injected polariton momentum $k$=1.3 µm$^{-1}$, the polariton velocity is $v_{flow} = \hbar k / m_{pol} = 4.6$ µm/ps ($m_{pol} = 2.0 \times 10^{-4} m_{electron}$). This velocity is higher than the speed of sound of the fluid for the polariton densities of our experiments, as evidenced by the presence of ship waves upstream from the obstacle in Fig. 2a (see Ref. 30). For this value of the momentum, we measure a value of the TE-TM splitting of 20 µeV, resulting in the effective magnetic field sketched in Fig. 1b.

Polaritons are injected in our microcavity with TM linear polarisation. This corresponds to a pseudospin pointing in the direction of the flow as marked by an arrow in Fig. 1a.

**Detection scheme.** The observations reported in this work require the complete knowledge of the spin of the polaritons. To do so, a complete polarisation tomography of the emission is realized through the measurement of the three Stokes parameters $S_1 = I_{TE} - I_{TM}$, $S_2 = I_D - I_A$ and $S_1 = I_{\sigma+} - I_{\sigma-}$, where $I_j$ is the light intensity emitted with polarisation $j$. This requires measuring six different polarisations: $I_H$, $I_V$, $I_{+45}$, $I_{-45}$, $I_{\sigma+}$ and $I_{\sigma-}$ which stand, respectively, for linear horizontal, linear vertical, linear diagonal, linear anti-diagonal, left circular and right circular polarised emitted intensity. A combination of waveplates and polarizing beam splitters is used to image each of the polarisation components of the emitted light on a CCD camera.

# Half-solitons in a polariton quantum fluid behave like magnetic monopoles

R. Hivet[1], H. Flayac[2], D. D. Solnyshkov[2], D. Tanese[3], T. Boulier[1], D. Andreoli[1], E. Giacobino[1], J. Bloch[3], A. Bramati[1*], G. Malpuech[2], A. Amo[3*]

[1] *Laboratoire Kastler Brossel, Université Pierre et Marie Curie, Ecole Normale Supérieure et CNRS, UPMC case 74, 4 place Jussieu, 75005 Paris, France*

[2] *Institut Pascal, PHOTON-N2, Clermont Université, University Blaise Pascal, CNRS, 24 avenue des Landais, 63177 Aubière cedex, France*

[3] *Laboratoire de Photonique et Nanostructures, CNRS, Route de Nozay, 91460 Marcoussis, France*

* e-mail address : bramati@spectro.jussieu.fr; alberto.amo@lpn.cnrs.fr .

## A.- CASE OF CIRCULARLY POLARISED PUMP

The mechanism to generate oblique half-solitons requires exciting the system with linearly polarised light[27]. As a proof of this statement, we show in this section that pumping with a circularly polarised beam, as performed in Ref. 28, leads to integer-like solitons which, contrary to half-solitons, are not affected by the effective magnetic field.

Supplementary Figure 1 shows the density, phase and pseudospin of a polariton fluid created with right circularly polarised ($\sigma_+$) light upstream from the defect. The corresponding initial pseudospin vector $\vec{S}$ therefore points in the *z* direction (along the growth axis of the sample, north pole of the Bloch sphere –Fig. 1a). The effective magnetic field $\vec{\Omega}_{TE-TM}$ induced by the TE-TM splitting lies on the plane of the microcavity (red arrow in Fig. 1b). The pseudospin and the magnetic field are then perpendicular to each other and this should lead to a precession of $\vec{S}$, even upstream from the defect. However, the density imbalance between $\sigma_+$ and $\sigma_-$ populations introduces an additionally energy splitting between these two components arising from the spin anisotropy of the polariton-polariton interaction. Polaritons with the same spin interact via the parameter $\alpha_1$ (see below), while polaritons of opposite spin present a much weaker interaction $\alpha_2 \approx -0.1\alpha_1$). This spin splitting gives rise to an extra effective magnetic field along the z axis (parallel to the injected pseudospin): $\vec{\Omega}_z = -S_3(\alpha_1 - \alpha_2)/2\, \vec{u}_z$. When we inject a large spin up population, this effective field is much higher than $\vec{\Omega}_{TE-TM}$, and no pseudospin precession takes place. Such behaviour is clearly visible in the Supplementary Fig. 1c, where the degree of circular polarisation is close to 1 and almost unperturbed close to the barrier. The strength of $\vec{\Omega}_z$ is however reduced far away from the defect where the density of polaritons decays due to their finite lifetime which allows the conversion of spin up to spin down polaritons (yellow regions).

Therefore, under strong $\sigma_+$ pumping, the $\sigma_-$ population remains weak everywhere, being completely determined by the transfer of particles from the other component. This is seen in Supplementary Figs. 1a and 1b, where one can observe integer (scalar) solitons in the $\sigma_+$ component, and their copy (but with a smaller density) in the other component. These integer solitons are not affected by the in-plane field $\vec{\Omega}_{TE-TM}$ and no acceleration is observed in that case.

The phase slices along the white dotted line shown in Supplementary Fig. 1e undergo a phase jump of approximately 0.7 $\pi$ in the $\sigma_+$ component (as well as in the $\sigma_-$, which just follows the dominant component) across each soliton, as expected for integer grey solitons.

Finally, the rotation of the linear polarisation angle is a specific signature of half-solitons, and it is strongly suppressed in this configuration as one can see in Supplementary Fig. 1c, representing the pseudospin, and in Supplementary Fig. 1f, which shows a slice of the polarisation angle along the white dotted line.

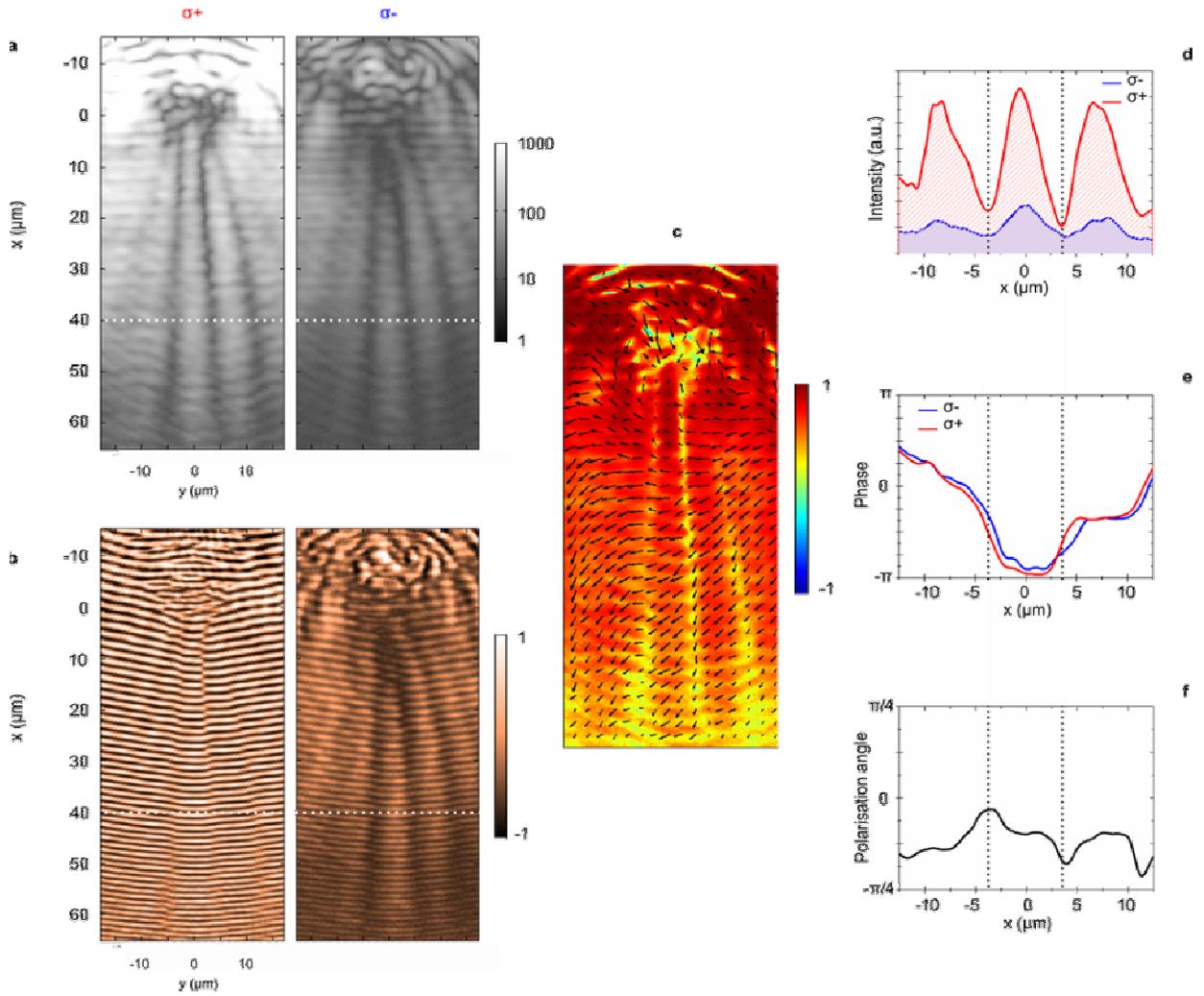

**Supplementary Figure 1 : Generation of dark solitons using a σ₊ polarised pump**. **a** Intensity of the light emitted by the microcavity in the σ₊ (left panel) and σ₋ (right panel) polarisation using a logarithmic scale. **b** Normalized fringes obtained from the interferences between the real space emission of the microcavity and a reference beam of constant phase. **c** Pseudospin map of the emission. The colours show the degree of circular polarisation and the arrows' direction the projection of the pseudospin on the linear polarisation plane (equator of the Bloch sphere). **d** Profiles extracted from **a** and along the white dotted line. **e** Corresponding profiles of the phase extracted from **b**. **f** Linear polarisation angle along the same dotted line.

## B.- NUMERICAL SIMULATIONS

We have performed numerical calculations reproducing in detail the experimental results shown in Figs. 1 to 4 in the main text. The result is shown in Supplementary Fig. 2. To accurately model the polariton fluid, we need to take into account its driven-dissipative character, that is, we should include the continuous pumping of polaritons in the excitation area and the losses arising from the polaritons escaping out of the cavity due to their finite lifetime. In order to do so we have used a set of modified Schrödinger and Gross-Pitaevskii equations for the spin-dependent photonic mean field $\phi_\pm(\vec{r},t)$ and the excitonic one $\chi_\pm(\vec{r},t)$ respectively, coupled via the light-matter interaction given by the Rabi splitting $\Omega_R = 5.1\,\text{meV}$:

$$i\hbar\frac{\partial \phi_\pm}{\partial t} = -\frac{\hbar^2}{2m_\phi}\Delta\phi_\pm + \Omega_R\chi_\pm + D\phi_\pm + \beta\left(\frac{\partial}{\partial x} \mp i\frac{\partial}{\partial y}\right)^2\phi_\mp + P_\pm e^{i(\vec{k}_P\cdot\vec{r}-\omega_P t)}f(\vec{r}) - \frac{i\hbar}{2\tau_\phi}\phi_\pm$$

$$i\hbar\frac{\partial \chi_\pm}{\partial t} = -\frac{\hbar^2}{2m_\chi}\Delta\chi_\pm + \Omega_R\phi_\pm + \left(\alpha_1|\chi_\pm|^2 + \alpha_2|\chi_\mp|^2\right)\chi_\pm - \frac{i\hbar}{2\tau_\chi}\chi_\pm$$

This set of equations is derived by minimizing the following Hamiltonian and adding pumping and dissipative terms treated via a dissipative formalism:

$$H = \sum_{\substack{\psi=\{\phi,\chi\}\\\sigma=\{+,-\}}}\left[\frac{\hbar^2}{2m_\psi}|\vec{\nabla}\psi_\sigma|^2\right] + \sum_{\sigma=\{+,-\}}\left[D|\phi_\pm|^2 + \frac{\Omega_R}{2}\left(\chi_\sigma^*\phi_\sigma + \chi_\sigma\phi_\sigma^*\right) + \frac{\alpha_1}{2}|\chi_\sigma|^4 + \frac{\alpha_2}{2}|\chi_\sigma|^2|\chi_{-\sigma}|^2\right] - \vec{S}\cdot\vec{\Omega}_{TE-TM}$$

The very last term is the magnetic energy of the condensate where $\vec{S}$ is the normalized pseudospin vector defined by:

$$S_0 = \sqrt{s_1^2 + s_2^2 + s_3^2}$$

$$\vec{S} = \begin{pmatrix}S_1\\S_2\\S_3\end{pmatrix} = \begin{pmatrix}\Re(\phi_+\phi_-^*)/S_0\\ \Im(\phi_+^*\phi_-)/S_0\\ (|\phi_+|^2 - |\phi_-|^2)/2S_0\end{pmatrix}$$

Here $s_j$ are the components of the pseudospin (without normalisation). $\vec{\Omega}_{TE-TM}$ is the effective field induced by the photonic TE-TM splitting given by:

$$\vec{\Omega}_{TE-TM} = \beta(k_x^2 + k_y^2)\begin{pmatrix}\cos(2\theta)\\ \sin(2\theta)\\ 0\end{pmatrix}$$

where $\theta$ is the polar angle. In the equations (1) and (2) we have used the transformations $k_{x,y} \leftrightarrow -i\partial_{x,y}$; $\tau_\phi = 20$ ps and $\tau_\chi = 400$ ps are the photonic and excitonic lifetimes respectively; $f(\vec{r})$ and $P_\pm$ are respectively the spatial extension and amplitudes of the *cw* pump of frequency $\omega_P = E_{LPB}(\vec{k}_P)/\hbar$ (we pump exactly on the bare polariton branch in this simulation) and wave vector $\vec{k}_P = 1.3 \mu m^{-1}$, acting on each spin component $\sigma_\pm = \pm$. The constant $\alpha_1 = 3E_b a_B^2/S$ (where $E_b = 10$ meV is the exciton binding energy, $a_B = 10^{-2} \mu m$ is its Bohr radius and $S$ is the normalization area) describes the intra-component exciton-exciton repulsion while $\alpha_2 = -0.1\alpha_1$ stands for the weak attractive inter-component interaction. $D(\vec{r})$ is a potential barrier that accounts for the photonic defect in the microcavity. The constant $\beta = \hbar^2 (1/m_\phi^{TE} - 1/m_\phi^{TM})/2$ is the strength of the effective magnetic field induced by the $k^2$-dependent photonic energy splitting between TE and TM eigenmodes[36]. We note that the excitonic TE-TM splitting is orders of magnitude smaller and is therefore neglected. Here $m_\phi^{TE} = 0.95 m_\phi^{TM}$ and $m_\phi^{TM} = 5 \times 10^{-5} m_0$ are the masses of the transverse electric and transverse magnetic cavity eigenmodes, and $m_0$ is the free electron mass. The photonic effective mass is taken as $m_\phi = 2(m_\phi^{TE} m_\phi^{TM})/(m_\phi^{TE} + m_\phi^{TM})$.

**Separation of oblique half-solitons with opposite charges**

Simulations based on the solutions of the above equations are shown in Supplementary Fig. 2. To illustrate the clearest way the action of the effective magnetic field on the trajectory of the half-solitons, and then to evidence their monopole behaviour, we have deliberately chosen a defect with a small radius (around 2 $\mu$m) to avoid the nucleation of multiplets of oblique solitons. Supplementary Fig. 2 reproduces all the panels of the Fig. 2, and shows a remarkable agreement with the experiment. In the simulation we can analyse the flow at larger distances than in the experiment, where the separation is larger.

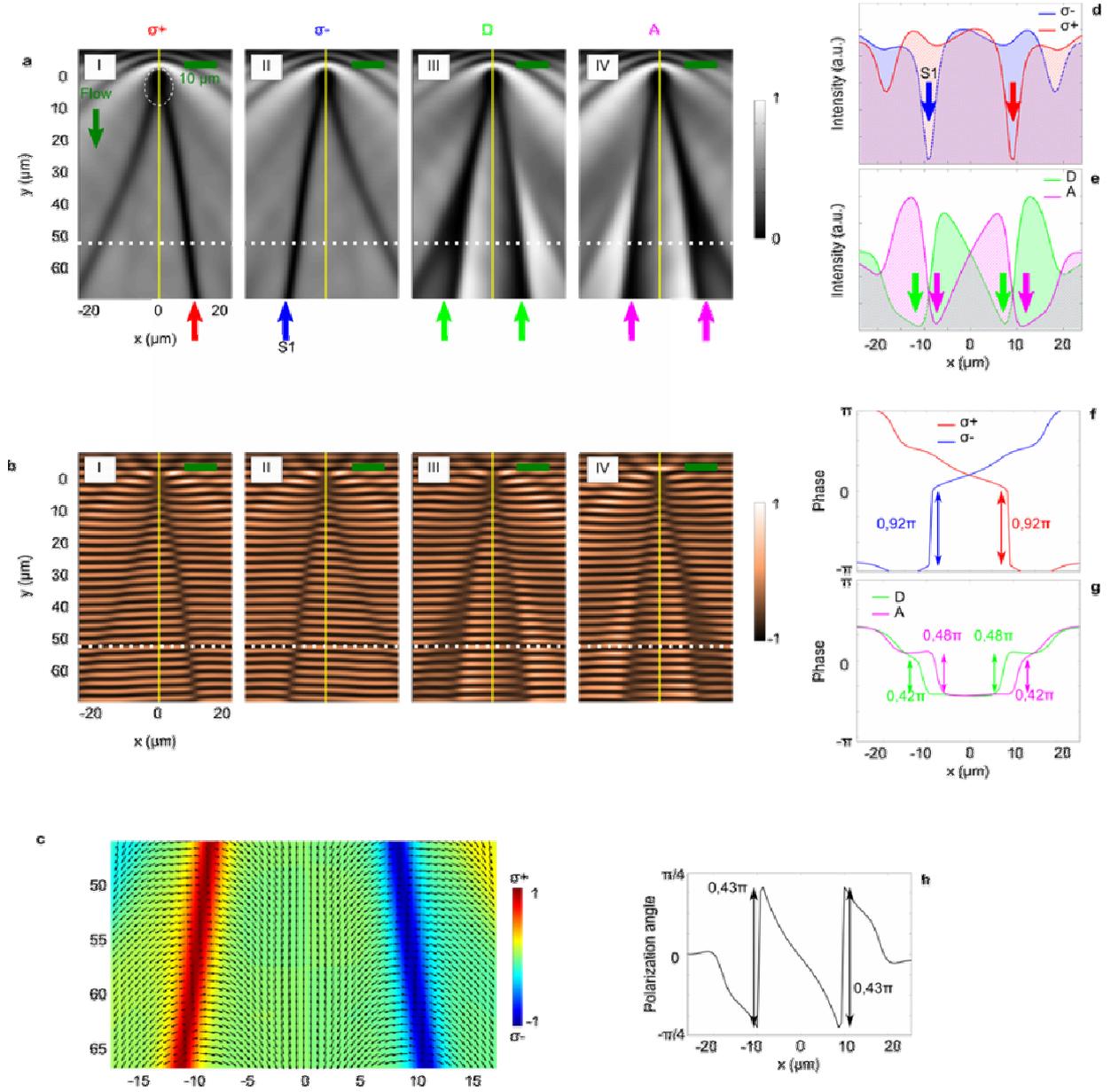

**Supplementary Figure 2.** Simulations showing the monopole-like behaviour of the oblique half-solitons (linearly polarised pump). **a** Calculated emission in the circular (I-II) and diagonal (III-IV) polarisations. **b** Corresponding interference patterns showing the local phase shifts of the fluid. **c** Calculated pseudospin extracted from the linear and circular degree of polarisation of the emission: the black arrows stand for the in plane projection of the pseudospin [$S_\parallel = (S_1, S_2)$, equator in the Bloch sphere] showing the inversion of the linear polarisation across the solitons (also visible in **h** showing a profile of the linear polarisation direction), and the colormap depicts the $S_3$ component evidencing the strong local degree of circular polarisation along the deep half-solitons. **c** and **e** Profile of the σ+ and σ- densities and their related phase, respectively, along the dotted line in **a** and **b**. **d** and **e** Density and phase profiles of the emission in diagonal polarisation.

Downstream and close to the obstacle, we see the formation of a pair of integer solitons. Each one of them can be considered as the composition of two half-solitons with equal contributions in the $\sigma_+$ and $\sigma_-$ components. As we get farther from the obstacle, $\vec{\Omega}_{TE-TM}$ accelerates each of the half-solitons in a direction depending on its charge (given by the $\sigma_\pm$ polarisation of their core). The action of this field results in the separation of the initially integer solitons into two sets of half-solitons, and it is the smoking gun of the monopole behaviour of half-solitons. This is clearly visible in panels I and II of Supplementary Fig. 2a, showing the emission from both circular components, as well as panels I and II of Supplementary Fig. 2b, showing the phase shifts undergone by the condensate.

If we concentrate on the left part of the flow (left side of the yellow line) we see that the soliton present in the $\sigma_-$ component [S1, visible in the panels 2a-II and 2b-II], initially overlaps with the $\sigma_+$ soliton, and it is bent towards the right at large downstream distances, becoming almost parallel to the flow direction. This displacement stabilizes the soliton, making it deeper. On the other hand, the soliton present on the $\sigma_+$ component has an opposite magnetic charge and is accelerated towards the left [panels 2a-I and 2b-I], with an orientation that tends to set the soliton trajectory perpendicular to the polariton flow. The $\sigma_+$ half-soliton thus gains speed with respect to the flow. As the soliton depth $n_s/n$ is related to its speed via the expression[29] $v_s = c_s\left(1 - n_s/n\right)^{1/2}$, where $c_s$ is the sound speed of the fluid, the $\sigma_+$ half-soliton soliton becomes shallower (as evidenced by the reduced phase shift visible in panel b-I) and, eventually, disappears. Note that the trajectories of the half-soliton are not linear but curved with a quadratic dependence with the distance from the defect which is a clear signature of their acceleration. The spatial separation between deep and shallow solitons is clearly visible in panel 2c showing a density profile along the white dotted line 52 $\mu$m away from the defect. The acceleration of the oblique half-solitons with opposite polarisation under the effect of the TE-TM effective magnetic field reveals the monopole-like behaviour of these objects. The sign of the magnetic charge associated to the half-solitons is given by the degree of circular polarisation of their core[12].

As obtained in the experiment and as originally predicted in Ref. 27, each oblique half-soliton also behaves as a domain wall with respect to linear polarisations, which appears most clearly on the diagonal (D)/anti-diagonal (A) polarisation basis defined by:

$$\begin{pmatrix} \psi_D \\ \psi_A \end{pmatrix} = \begin{pmatrix} \cos(\pi/4) & \sin(\pi/4) \\ -\sin(\pi/4) & \cos(\pi/4) \end{pmatrix} \begin{pmatrix} \psi_x \\ \psi_y \end{pmatrix}$$

$$\begin{pmatrix} \psi_x \\ \psi_y \end{pmatrix} = \frac{1}{\sqrt{2}} \begin{pmatrix} 1 & 1 \\ i & -i \end{pmatrix} \begin{pmatrix} \psi_+ \\ \psi_- \end{pmatrix}$$

The corresponding photonic emission intensities are shown in the panels 2a-III and 2a-IV and the respectives phases in the panels 2b-III and 2b-IV. The domain wall behaviour is underlined in the density profiles of the panel 2d.

Finally the half-integer character of the oblique solitons is demonstrated in the panels 2f and 2g, where the calculated phase profiles along the white dotted line are displayed. As one can see, the phase jumps through the deepest half-solitons on the diagonal linear basis (panel 2g) are one half of the one obtained on the circular polarisation basis (panel 2f). As well, the polarisation changes from diagonal to anti-diagonal, which means a rotation of its direction of $\pi/2$ across the deep solitons as it can be seen in the panels 2c and 2h.

**Supplementary references**